\documentclass[sigconf]{acmart}

\usepackage[]{mdframed}
\usepackage[font=small,skip=5pt]{caption}
\usepackage{enumitem}

\AtBeginDocument{%
  \providecommand\BibTeX{{%
    \normalfont B\kern-0.5em{\scshape i\kern-0.25em b}\kern-0.8em\TeX}}}

\setcopyright{acmcopyright}
\copyrightyear{2018}
\acmYear{2018}
\acmDOI{XXXXXXX.XXXXXXX}

\acmConference[Conference acronym 'XX]{Make sure to enter the correct
  conference title from your rights confirmation emai}{June 03--05,
  2018}{Woodstock, NY}
\acmPrice{15.00}
\acmISBN{978-1-4503-XXXX-X/18/06}

\begin{document}

\title{Using Peer-Customers to Scalably Pair Student Teams with Customers for Hands-on Curriculum Final Projects}

\author{Edward Jay Wang}
\email{ejaywang@ucsd.edu}
\affiliation{%
  \institution{University of California San Diego, The Design Lab}
  \city{La Jolla}
  \state{CA}
  \country{USA}
}
\renewcommand{\shortauthors}{Anon}

\begin{abstract}
  Peer-customer is a  mechanism to pair student teams with customers in hands-on curriculum courses. Each student pitches a problem they want someone else in the class to solve for them. The use of peer-customers provides practical and scalable access for students to work with a customer on a real-world need for their final project. The peer-customer, despite being a student in the class, do not work on the project with the team. This dissociation forces a student team to practice customer needs assessment, testing, and surveying that can often be lacking in self-ideated final projects that do not have resources to curate external customers like in capstone courses. We prototyped the use of peer-customers in an introductory physical prototyping course focused on basic embedded systems design and python programming. In this paper, we present a practical guide on how best to use peer-customers, supported by key observations made during two separate offerings of the course with a total of N=64 students (N=29 Y1 and N=35 Y2).  
\end{abstract}




\maketitle

\section{Introduction}
In undergraduate design-oriented courses, team projects is a common mode of learning and often takes the form of hands-on final projects. Whereby a team is formed through mutual interest, friend groups, or through instructional team pairing to promote diverse teams \cite{warhuus_i_2017, linsey_collaborating_2008, striker_scalable_2020, spoelstra_toward_2014, hong_using_1995}. The team then formulates an idea that is suitable for the technical skills learned in the class (e.g. mobile programming, embedded devices, VR/AR, computer vision, etc). This mode of project team construction is scalable as student teams come up with their own projects \cite{lee_idea_2019, sedaghat_factors_2018}. 

However, students working on ideas born out of their own ideation has the risk of establishing counterproductive habits early on in their career development. Because the team came up with the idea, they are themselves inherently the “customer.” Thus, in the course of development, the student team will design and iterate on solutions based on what they themselves (1) likes and dislikes and (2) can mold an idea to the skills they have. This is much more akin to DIY projects rather than designing for a customer. In DIY projects, the only judge of whether a solution works is based on personal taste. If it fits my hands, then the design is right. If I like the UI, then the design is right. 

In comparison, in senior design projects, often called capstone projects, student teams are formed by instructional teams and paired with real-world customers \cite{allen_experiential_2021, paasivaara_collaborating_2019,paasivaara_how_2018, isomottonen_value_2008}. These real-world customers are curated by the instructional team through contact with industry relationships and on-campus organizations (i.e. clubs, facilities, research labs). External customers in these capstone projects provide a number of benefits to student education: 

\begin{itemize}[leftmargin=*]
  \item \textbf{Customer provided constraints} – The customers’ needs are what defines the specifications that the solution must meet \cite{bond_difficult_1995}. 
  \item \textbf{Technical solution has to meet the need, not the other way around} – Because the customers have specific problem to be addressed, the solution that is developed has to mold to their needs. That means students often have to learn new technical skills they may not have possessed already to build the solution needed. 
  \item \textbf{Customers have to like the solution} – The end goal is to satisfy the customer, not the instructor. The instructor is there to help you meet the customer’s needs and assess the approach and degree the solution meets the customer’s needs.
  \item \textbf{Objective system tests become essential} – Because the customer is not part of the team and can only provide feedback during scheduled times, it is important to define metrics that the solution has to meet and test the system against these metrics. 
\end{itemize}

The use of external customers prepares students towards working with real-world clients. However, to incorporate external customers into non-capstone project courses is challenging. Instructional teams have to acquire and liaison a pool of customers commensurate to the size of the class. Section from section can have different class sizes. Engagement from community can fluctuate seasonally and year-by-year. Dramatically increasing instructor burden. 

\section{Peer-Customers}
\begin{figure}[!h]
    \centering
    \includegraphics[width=1\columnwidth]{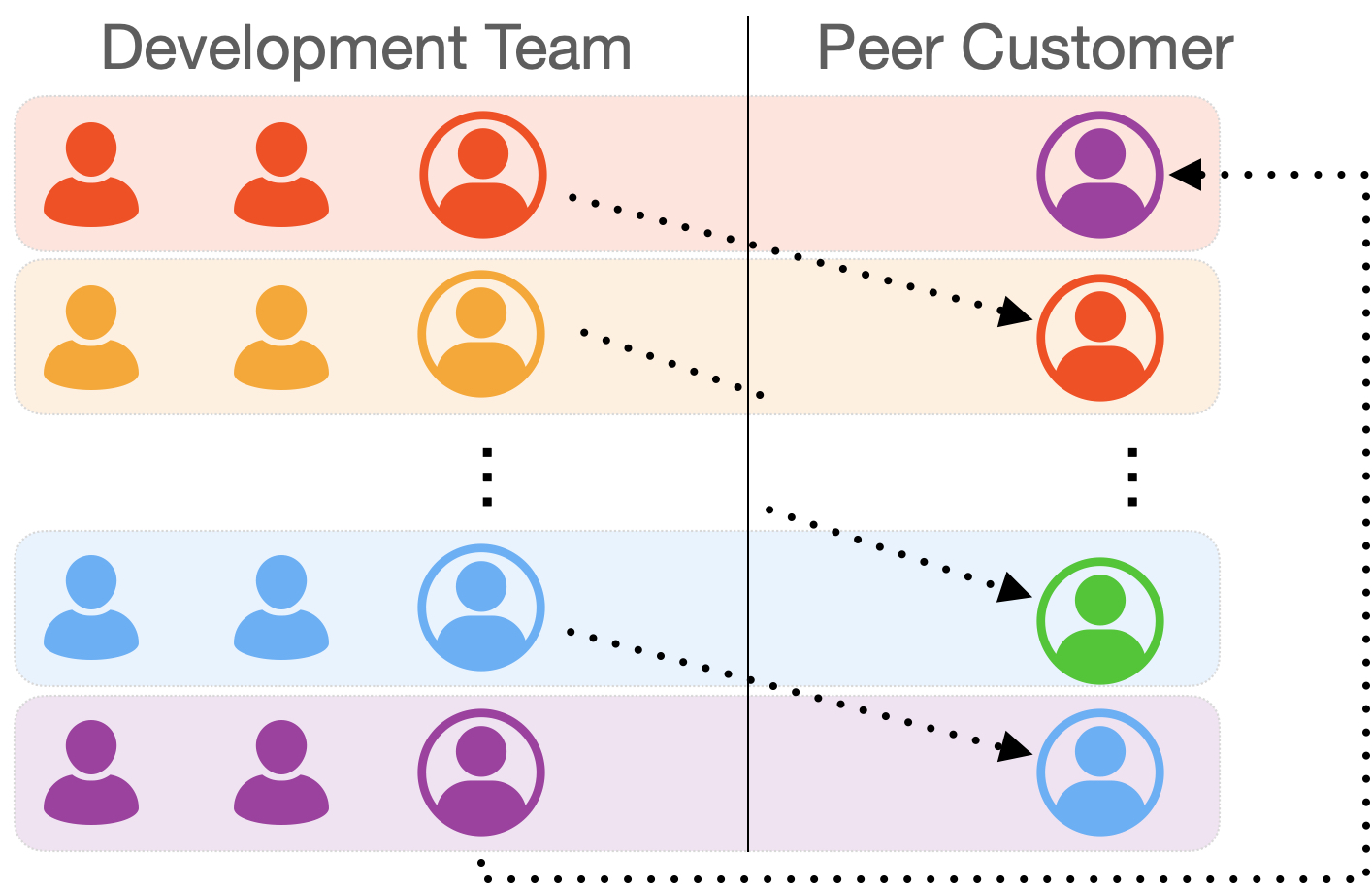}
    \caption{ Schema of development team and customer pairing.}
    \label{customer_pairing}
\end{figure}

In our class, ECE[Class Name Anonymized for Review], we prototyped a mechanism to provide each student team a customer who is not part of the student team, while doing so without increasing instructor burden as the class size grows. \underline{\textbf{Peer-Customers}} is a form of customer pairing where students within the class will pitch one problem that they wish someone in the class would make a solution to solve for them. A portion of the ideas are then chosen based on a mix of student voting and instructor curation. Once chosen, the teams are split such that each team has one student who is a peer-customer. However, this student is NOT the customer for their team. Instead, they are the peer-customer for another team. In this way, every student development team is paired with a peer-customer who is in the class, but not in the development team. Figure \ref{customer_pairing} visualizes this pairing scheme. 

The benefit of leveraging Peer-Customers is that it simulates the effect of having an external customer. Although not a perfect replica of a real-world paying customer, peer-customers provide real-world problems someone actually wants solved. With the Peer-Customer being part of the class, it is possible to schedule mandatory in-class activities for meeting with the customer. This mechanism is also robust to fluctuating class sizes as the number of peer-customers is directly proportional to the number of enrolled students. 

\textbf{\emph{Educational Context:}} ECE[Anon] is a course that teaches students basic prototyping with electronics, microcontrollers, and python programming for signal processing. Students enrolled in the class are required to have taken a course equivalent to introduction to programming. The enrollment is ~30 people each section with one instructor and two 10-hour TA allocation. The class meets for one 3-hour lecture/in-class activities session and one 3-hour lab session for open work with TA availability. Only the lecture/in-class activities session is mandatory attendance. We have implemented Peer-Customers in two section offerings, Winter 2022 and Spring 2023, with enrollment of N=29 and N=35 respectively.

\subsection{Customer Pitches}
As part of a graded in-class activity, each student pitches a \textbf{need}. A need is defined as something they personally experience and would like solved. What is important is that this need is personal and cannot be something they project someone else might want. Each student prepares a 1-slide, 1 minute presentation where they describe their need to the class. 

\begin{figure}[!htpb]
    \centering
    \includegraphics[width=1\columnwidth]{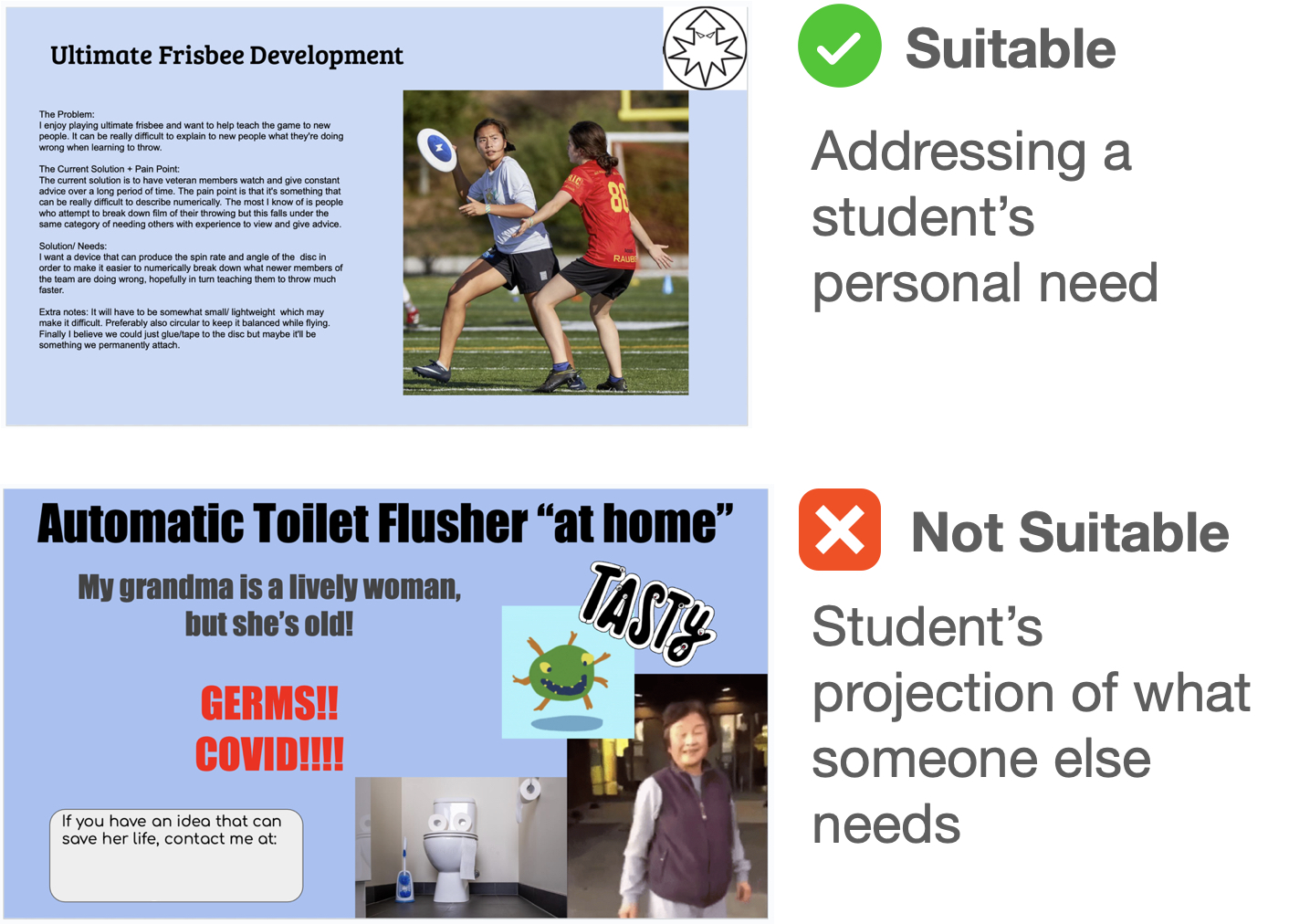}
    \caption{Examples of suitable and not suitable customer pitches}
    \label{customer_pitch}
\end{figure}
\setlength{\belowcaptionskip}{-20pt}

Figure \ref{customer_pitch} shows example pitches from students in the class. The top example is a suitable pitch about ultimate frisbee training where the student is a player on the ultimate frisbee team. They are a senior team member who trains other students and are looking for a tool that can help him during his coaching of new players. He wants a frisbee that can track and show the speed of the rotation, a key metric on how stable the flight will be. The bottom example is a not suitable pitch about a smart toilet that can automatically flush for the student's grandmother who has mobility issues. Because the need is a projection of what the student thinks his grandmother would benefit from having, it is not suitable. The student cannot provide meaningful assessment of how easy it is to use the system as they themselves do not share the same mobility issues. At best, they can act as a go-between with the real customer, his grandmother.

\subsection{Customer Selection}
After all students have pitched their needs, an online survey is shared with the class where students get to vote on their top-3 favorite needs. The instructor will assess whether a need is (1) addressable with the technical skills the students will either directly learn through the homework assignments or are prepared to pick-up based on the learning objectives of the class and (2) an interesting problem that would result in a fulfilling project. The student vote provides some additional feedback to the instructor for use when choosing between two equally suitable ideas. 

The number of needs selected is proportional to the class size and the number of students to be assigned to each team. For example, in a class of 35 people with a desired team size of 3 students, 11 peer-customers will be chosen. The 2 extra students are then assigned to projects that the instructor expects might be more difficult, forming nine 3-people teams and two 4-people teams. 
\vspace*{-5mm}
\setlength{\belowcaptionskip}{-10pt}

\subsection{Customer Pairing}
Teams are then formed by the instructor in the following way:
\begin{itemize}
    \item Each team is assigned a peer-customer
    \item Each team is assigned 3 students to the development team. 
    \item 1 of the 3 students is a peer-customer for another team. 
\end{itemize}

The key here is that the peer-customer is NOT part of the development team. Their involvement with the team is only during customer meetings to be interviewed and to provide feedback on the solution, but are not supposed to engage in the active making of the solution. 

Note that each development team should only have one student who will act as a peer-customer. This is to facilitate in-class activities where the peer-customer will leave their development team and meet with the team they are the customer for. This way, teams will have enough none-customer members available for working with a peer-customer. 

\section{Class Activities for Customer Engagement}
The development team will treat their customer as a client, who they need to understand the needs of and design a prototype of a solution that aims to address their problem. To do so, the team will engage with the customer through various asynchronous and synchronous ways. A number of classroom activities are structured to effectively utilize peer-customers during scheduled class meetings to reduce added burden on the students acting as a peer-customer. 

\subsection{In-class customer meetings}
\begin{figure}[!htpb]
    \centering
    \includegraphics[width=1\columnwidth]{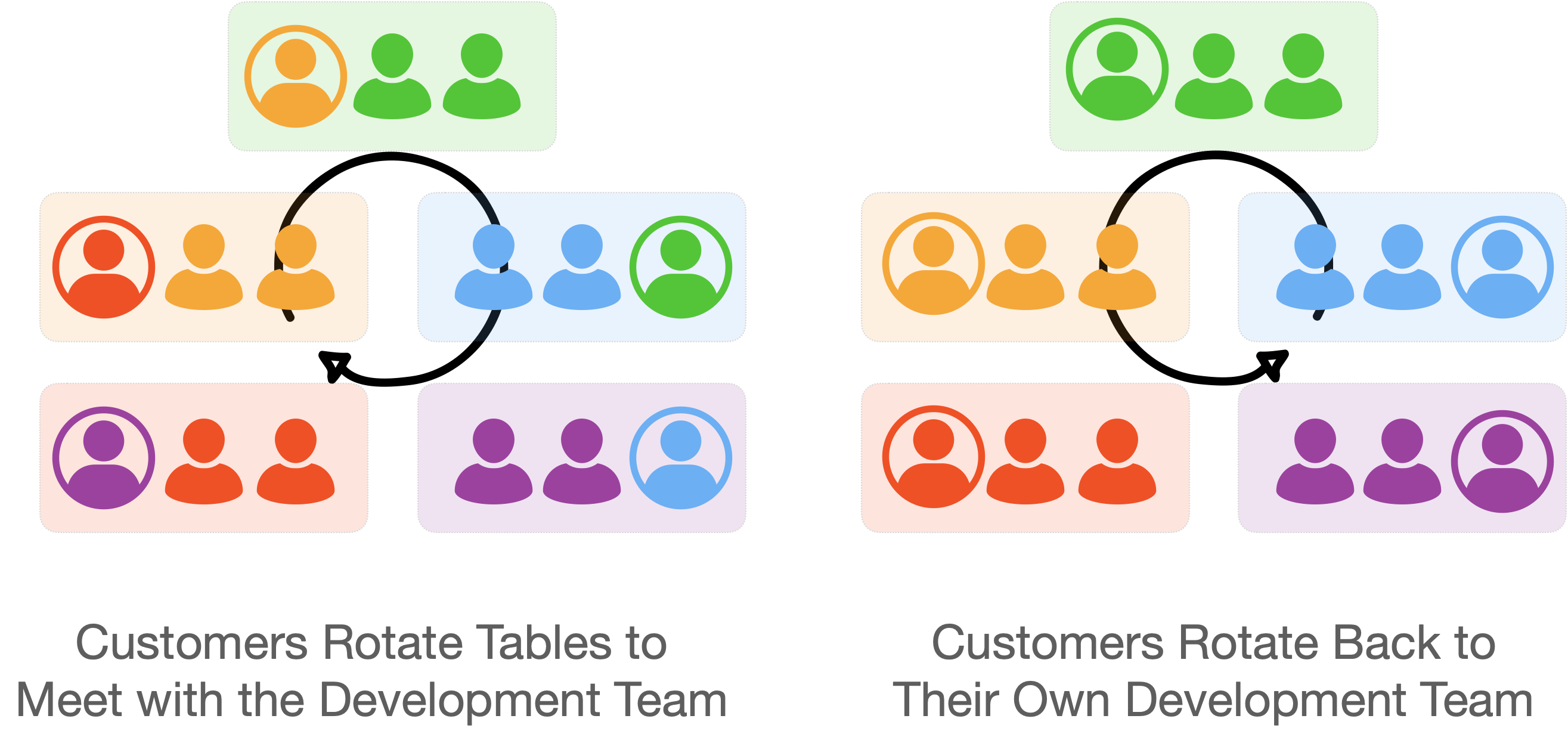}
    \caption{In class activity allows the development team to meet with their peer-customer. The peer-customer simply rotate by one table and rotate back to their table at the end of the class activity.}
    \label{class_rotation}
\end{figure}
\setlength{\belowcaptionskip}{-10pt}
A unique feature of the peer-customer mechanism is that the customer can be expected to appear to class meetings as they are themselves students enrolled in the class through mandatory in-class activities that are part of the grade (participation score). 

In-class activities are suitable for when the development team needs to meet with their customer, such as for the initial customer meeting, needs assessment, customer check-in, and prototype demonstrations. By having teams sit together and ordered by team numbers, peer-customers can easily rotate one table over to meet with the team. With three people teams, that means two members will stay behind to meet with the peer-customer. After the meeting, the peer-customer rotate back to their own development team and receives a debrief from their team of what they learned from their customer. 

\begin{figure}[!htpb]
    \centering
    \includegraphics[width=1\columnwidth]{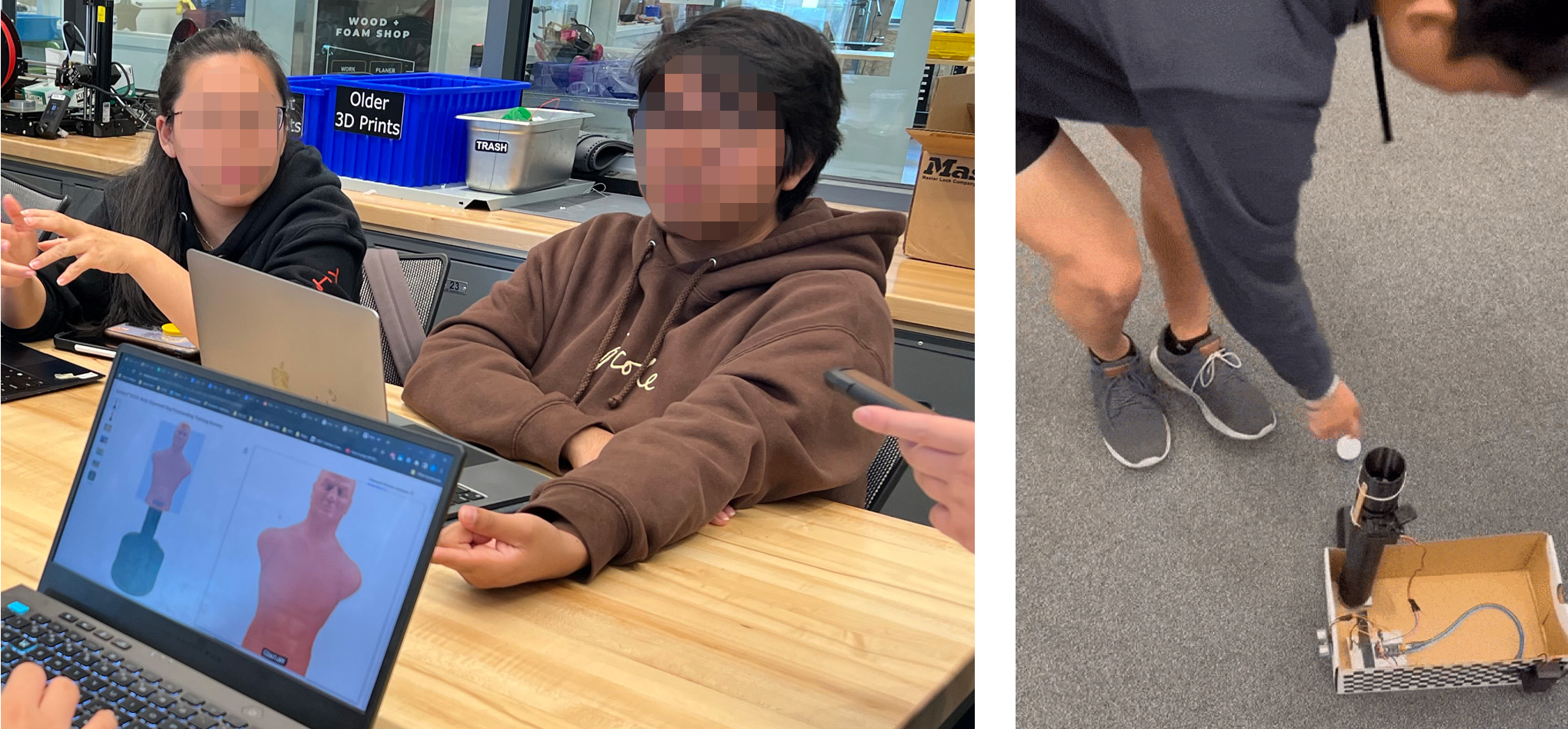}
    \caption{Left: Customer showing the team what the form factor they need. Right: Customer testing the prototype.}
    \label{class_rotation}
\end{figure}
\setlength{\belowcaptionskip}{-10pt}
\subsection{In-class presentations}
Development teams prepare three in-class presentations: (1) initial solution proposal, (2) progress update, and (3) final presentation.  In each of these presentations, the development team presents for 5-7 minutes, updating the entire class on their project. These presentations mimic progress update presentations a team would typically schedule and deliver for their customer in a capstone project. Having these presentations in class allows for the customer to see the presentation to get an update without having to schedule outside meeting times with the team. 

\subsection{Customer Survey}
Not all activities needs to be completed synchronously. For example, after the initial customer meeting and needs assessment, the development team are asked to create a list of 5 features that they want to assess the customer's requirement using the Kano Method. This is formed as a survey and sent to the customer electronically. 

\subsection{Sample Course Schedule}
The following outlines a 10-week quarter course structure that integrates the class project schedule with the technical learning components.
\begin{figure}[!hb]
    \centering
    \includegraphics[width=1\columnwidth]{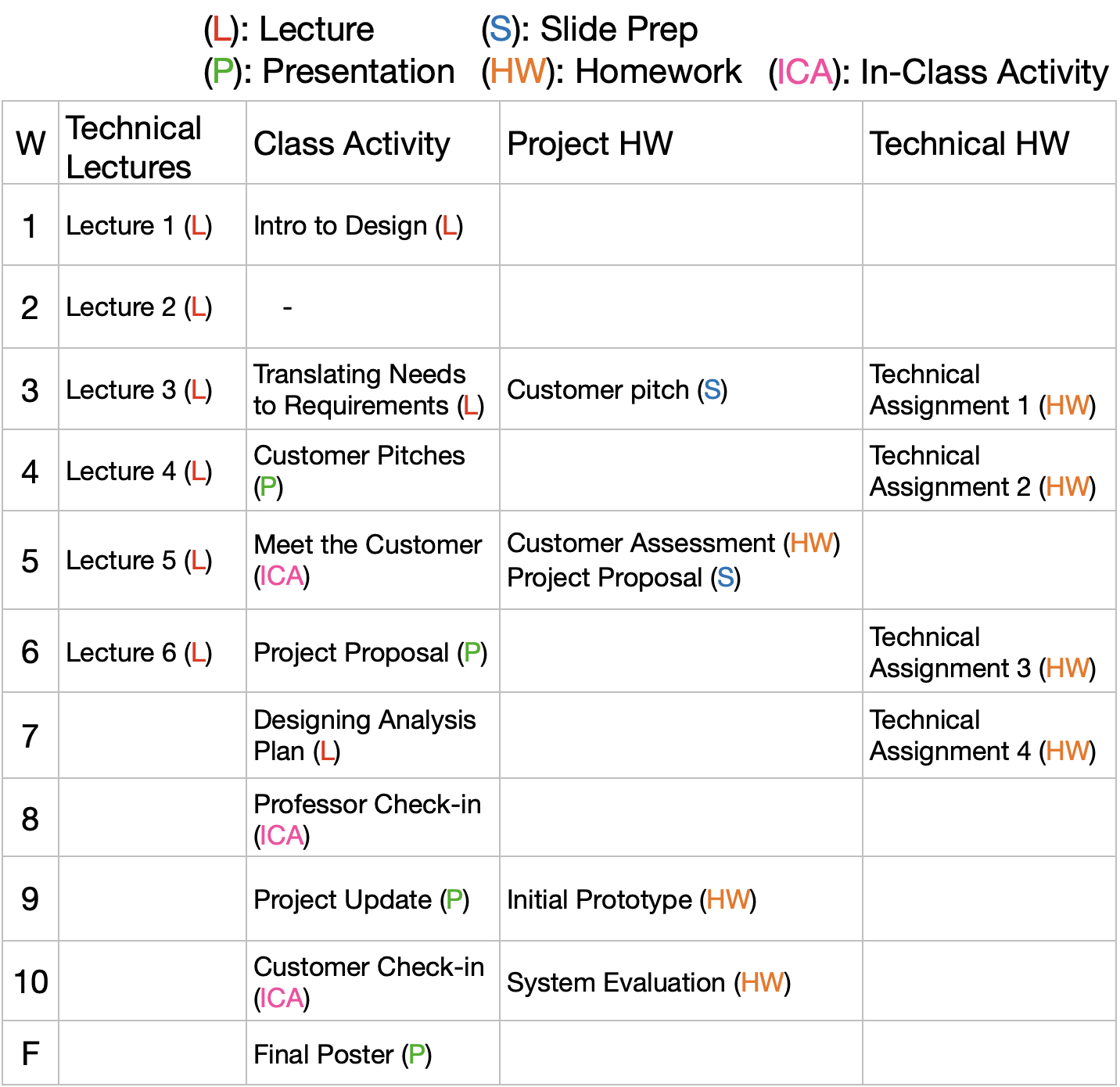}
    \caption{Sample Schedule of a 10-week Quarter}
    \label{schedule}
\end{figure}
\setlength{\belowcaptionskip}{-10pt}

 In our class, there are four technical assignments. In addition, there are three design assignments that are related to the project itself: (1) initial needs assessment, (2) solution formation and development, (3) solution testing and customer feedback. An important factor that ensure there is enough time between the initial needs assessment to solution development is to make sure the customer pitches occur early on in the quarter. As such, we interweaved technical assignments with design activities/assignments throughout the quarter. 

\section{Case Studies and Key Observations}
\subsection{What The Customer Wants Is What Matters, Not What the Professor Wants}
One of the most powerful effect that the use of peer-customer provides  is that the development team 
\textbf{must} prioritize the customer's needs. Although this may seem obvious, the degree to which this was valuable became abundantly clear to our instructional team in a highly memorable interaction with a team. The following is a case study where our team has since used to help motivate the concept of peer-customers to both other instructors and students on the value of working with a customer: 
\vspace{10pt}

\begin{mdframed}
\emph{\textbf{Case Study: "Light-up Fidget Cube"}}
The customer's pitch amounted to wanting a large fidget cube that lights up like a disco ball. As the team was developing the solution, they came to the instructor with a concern. In their latest customer check-in, they proposed a number of design alternatives for how the cube will light up. In their design, the buttons, nobs, and joystick on the fidget cube will control the lighting patterns. Dimming, flickering, and changing colors based on how you clicked it. However, their customer didn't like that. Instead, they don't want any of the buttons and nobs to control the light. What they want is for the light to "randomly do it's thing" and the only control they want is that the cube only lights up when they pick it up from the table, but should turn off when its just sitting there. 

That meant that the team did not need to wire any of the buttons and nobs and simply put them in for tactile feedback. The team was worried that the project would be too simple and that they will receive a bad grade, and asked the instructor if they should instead make the cube light up with input from the buttons. 

The instructor told the students, "The only thing that will end up with a bad grade is if you don't give your customer what they want. If they don't want it to control the lights, well great! Less work for you. Now figure out how to detect when they pick up the fidget cube."
\end{mdframed}

\vspace{10pt}
Although on the surface, it may seem like the customer's request is a bit arbitrary, but upon further examination, it makes perfect sense. Fidget cubes are meant to be an ambient distraction that should not occupy cognitive load. In fact, by playing with a fidget cube, it helps one focus on the actual task at hand, perhaps watching a lecture. But if the fidget cube instead acts as a lighting controller, it demands the attention of the user and becomes counter productive.  

This is a very interesting situation that would not have happened in a traditional final project where the students ideated their own project. This need to actually solve someone's needs conflicts with the desire to cram everything the student has learned in a class. What is particularly then powerful with the use of a peer-customer is that there is only one key requirement. \textbf{Satisfy the customer's need}. The instructor's job then is to (1) remind the team that their goal is to satisfy the customer, (2) help them understand what might be the reason the customer has a specific need, and (3) brainstorm with them how to pivot/redesign to best address the needs of the customer. 

\subsection{Using Design Probes to Gain Customer Feedback} 
The case study of the light-up fidget cube also provides us with another important aspect around aligning the solution to the customer's needs. Because the development team can't read the customer's mind, they have to come up with ways to truly understand them. This is where design probes become important. Design probes are sometimes simple mockups or samples that helps to illustrate a concept. Giving something physical to the customer such that they can try it and give feedback.

Another aspect that was important to the customer in our case study was the size of the cube. They wanted something "large," but that is quite a subjective measure. The team used a set of simple taped together blocks of wood as an initial design probe to quickly assess which one the customer liked the best. The customer, having played with the wooden blocks, noted that the sharp edges were unpleasant. With that in mind, the team 3D printed multiple cubes with varying curved edges for the customer to choose from. Finally, in their final iteration, the customer noted that their initial design only has the LEDs peeking out of the solid cube, but they want the whole cube to light up. The instructor then helped the team print the design with clear resin which allowed for the entire cube to light up. 
\begin{figure}[!htpb]
    \centering
    \includegraphics[width=1\columnwidth]{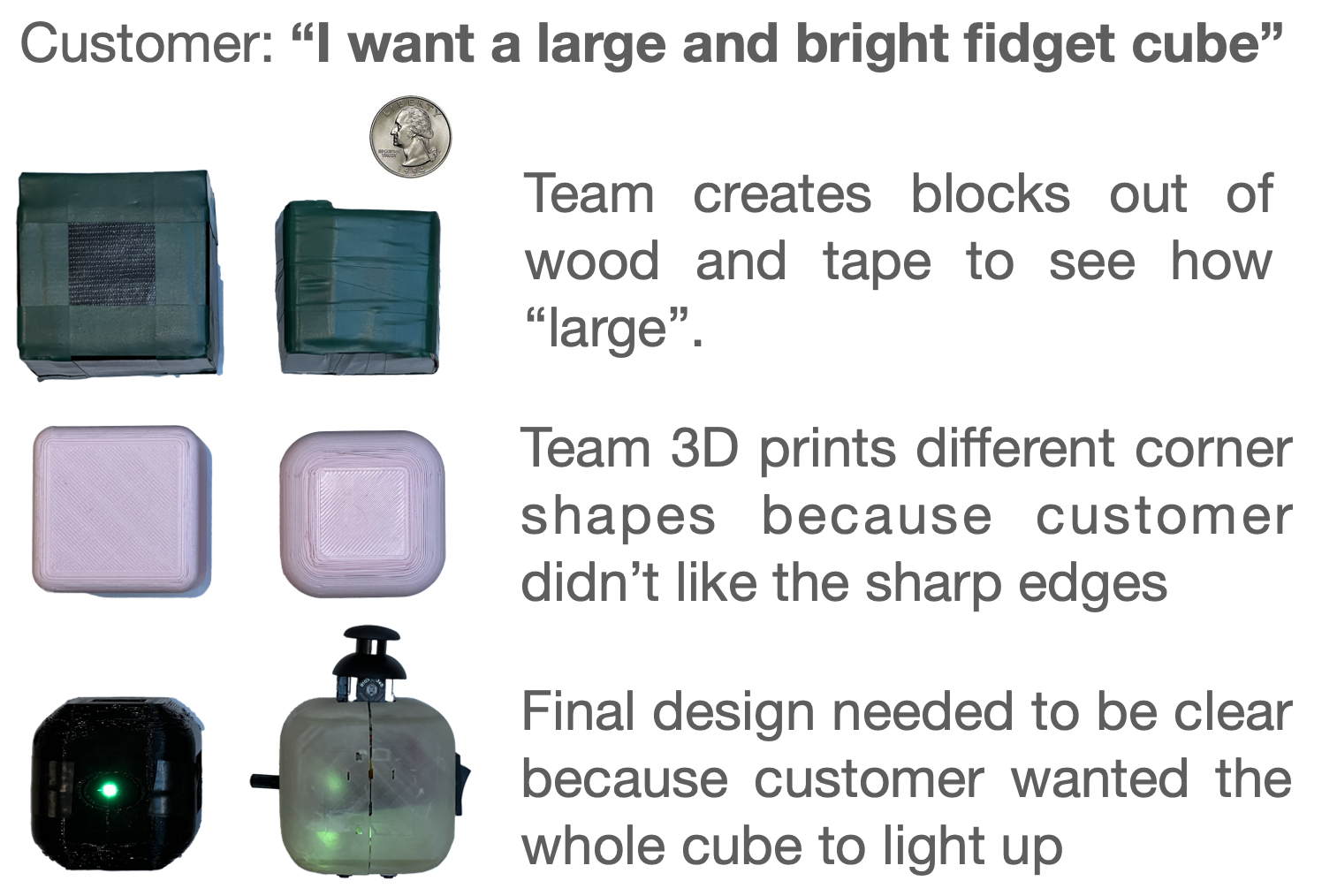}
    \caption{Example of how the development team needs to develop design probes to understand the customer's initial need and subsequent customer feedback shaped the direction of the design.}
    \label{iterating}
\end{figure}

The use of the design probe here, though simple, was instrumental for the team to actually understand the customer. This iterative process would not have happened in a self-ideated project because the team would have been able to know how big of a cube they personally liked, and simply went with that design. Thus missing out on learning how to use design probes effectively. 

\subsection{Solution Development not Product Development}
An important distinction is that the use of customers in a course means that the teams is operating as a client and design agency relationship. That is distinct from product discovery that is meant to be used by a pool of potential customers. This should be considered when leveraging peer-customers in an educational setting. If the course focuses deeply on customer discovery, need finding, and new product ideation/development, the use of peer-customers may not be appropriate. This is because the customer is coming with their needs and wants a solution tailored. In this way, the course will focused more on customer needs assessment, background research, and solution development. 

If a blend of customer discovery is desired within the learning objective, it is possible to add an additional background research portion to the project. In our class, we asked the students to use the Kano Analysis method to assess the requirements of the customer. The Kano Analysis \cite{xu_analytical_2009} uses three questions to assess if (1) How would one feel if the feature is included, (2) How would one feel if the feature is NOT included, and (3) How important is the feature. The Kano Analysis creates insight into even a single customer's desire for a feature. In addition to acquiring the requirement of their customer, we additionally ask the team to use the same survey on a broader audience to assess the generalizability of this tailored solution. To target a relevant audience, the team asks the customer to distribute their survey to the customer's network who share the same interest/need. 

\vspace{10pt}

\begin{mdframed}
\emph{\textbf{Case Study: "S-K-A-T-E Score Board"}}
The customer explains that skateboarders play a game called SKATE, which is reminiscent of the game of HORSE in basketball. Whereby one skateboarder demonstrates a trick and the other skateboarder subsequently attempts to replicate the trick. If the replicating skateboarder misses the trick, they accumulate one letter from the word SKATE. The one demonstrating the trick rotates after each trick. The first to accumulate all five letters of SKATE loses. This game can be played with 2 players, but can also be played by a bigger group where everyone replicates the trick. 

The team found that both the customer and in the larger survey (N=10), features that everyone shared is the ability to increment/decrement scores individually, reset the game, and display the current SKATE letters for each player on a screen. However, for the larger group, features such as LEDs/Buzzers for indicating someone lost and being able to track more than 2 players is attractive. While for the customer, they only want the screen without buzzers and only care to track 2 players. 

Based on these information the team chose to focus on the 2 player version of the design as that is what their customer wants and reducing complexity of the system to only including a screen.
\end{mdframed}

\vspace{10pt}
This case study captures a good example of how a solution is desirable for a bigger audience, with the customer wanting a specific feature that they want within the entire design space. 

An important aspect to include in the lecture/assignment instruction is to position the reasoning for doing this broader survey. It is important for the team to understand that their goal in doing this survey is to have a better understanding of what other potential customers with a general shared interest might like. But it is not to over-shadow the requirement of their customer. This is akin to how some contract design/engineering firms operate, where individual contracts needs to meet the exact specification of the customer (a custom solution). But the firm might assess how generally applicable this solution is to the broader audience and know what is a special need of their customer, and what might be features that are shared by a bigger audience. 

\subsection{Project Fit}
One of the main concerns at the initial conception of the use of peer-customers is around whether the proposed needs by students in the class will be appropriate. The following are some themes we noticed are key considerations that is important for successful gathering of a set of good projects.

\subsubsection{Need is too difficult to address}
Some concerns brought up by students of the class in the first iteration revolved around worries that the problems proposed might be too difficult or that the right solution might require skills beyond the scope of the class. What we found is that most of the problems that would be on the difficult scale ended up being good projects because they typically provided more open-ended solution spaces and there were usually ways to create scaled down solutions that would make it possible to create a version one prototype. 

We, however, did not find too many need proposals that required skills that is far outside the scope of the class. Our class is focused on hardware software interfaces with the use of Arduino, Bluetooth, signal processing, and physical prototyping. Because the students know what class they are taking, very few students proposed needs that clearly was unaddressable by the skills they would be learning in the class. Out of two offerings we implemented peer-customers, which included a total of N=64 (N=29 in Year1 and N=35 in Year2) pitches, we found only N=8 pitches (N=4 in Year 1 and N=4 in Year 2) that would be inappropriate for a class project. These needs typically had no physical component, and mostly required pure software implementations that would be pure cloud computing or webapp systems. Given that we only need 1/3 of the pitches to be selected, this did not pose a big issue. 

\subsubsection{Solved by existing products}
A common issue that emerged is the request for a solution that would otherwise be addressed by an existing product, but the student thinks the existing product is too expensive. These do not make for good projects. In fact, this is a very common misconception in engineering courses where students create hackathon-level prototypes and claim that one of the key advantages to their "new" solution is better than the product because the arduino is cheaper. This does not take into account the cost of design, manufacture tooling, packaging, marketing, and more that is involved in the creation of a product. 

Perhaps one of the most common offender of this is the request for a remote light switch. Out of the N=64 pitches, N=7 pitches were about wanting an automated/remote controlled light switch. Many citing existing solutions being too expensive. This is grossly untrue given the sheer abundance of existing consumer solutions readily available. 

In future iterations of the class, an improvement around customer pitching instructions will include information that excludes pitching needs purely based on cost concerns. This will be motivated by a discussion around the real cost involved in a product. 

\subsubsection{Crowd favorite vs instructor choice}
In the first iteration of the course, instead of the instructor choosing the peer-customers, the peer-customers were chosen based on popular vote. However, this was removed in the second iteration. There were a few issues that we observed in the popular vote mechanism:
\begin{enumerate}[leftmargin=*]
    \item There were a lack of diversity in the ideas chosen. Most of the ideas that gathered the most votes had to do with "too lazy to get out of my chair, I want a remote to do X" or "I can't wake up in the morning, I want my alarm to do X". 
    \item More unique ideas that were particularly personal, such as a less common hobby like motorcycling, off-road 4x4 driving, were left out of the top rankings. 
    \item Some ideas chosen by the class were inappropriate for the scope of the class due to them being pure software problems, as mentioned previously. 
\end{enumerate}

We ultimately chose to deviate from the popular vote mechanism even in the first iteration and replaced a few peer-customers after the initial vote, but still largely followed it. In the second iteration, we only used the votes as a rough guide, but mainly chose based on instructor discretion of what would likely lead to the most interesting project. Although the voting result is only loosely used, we still believe it provides some value to see if there are some ideas that clearly garnered a large number of votes but perhaps don't resonate with the teaching team. 

\subsection{Ensuring Participation}
Particular focus was made to make sure the peer-customer attended class as they were needed for the customer interviews. We leveraged a few mechanisms that attempted to increase class attendance.

\subsubsection{Incentive for Peer-Customers to Participate}
We further incentivised students to pitch interesting problems to be selected as peer-customers by providing a grade incentive. A peer-customer receives a 50\% weighting on their lowest technical assignment grade. With each technical assignments being worth 10\% of the total grade, this led to a significant incentive to pitch a good project. In future iterations of the class, we plan to make this incentive contingent on being present for ICAs to make sure that peer-customers continue to engage to earn this incentive. 
\begin{figure}[!htbp]
    \centering
    \includegraphics[width=1\columnwidth]{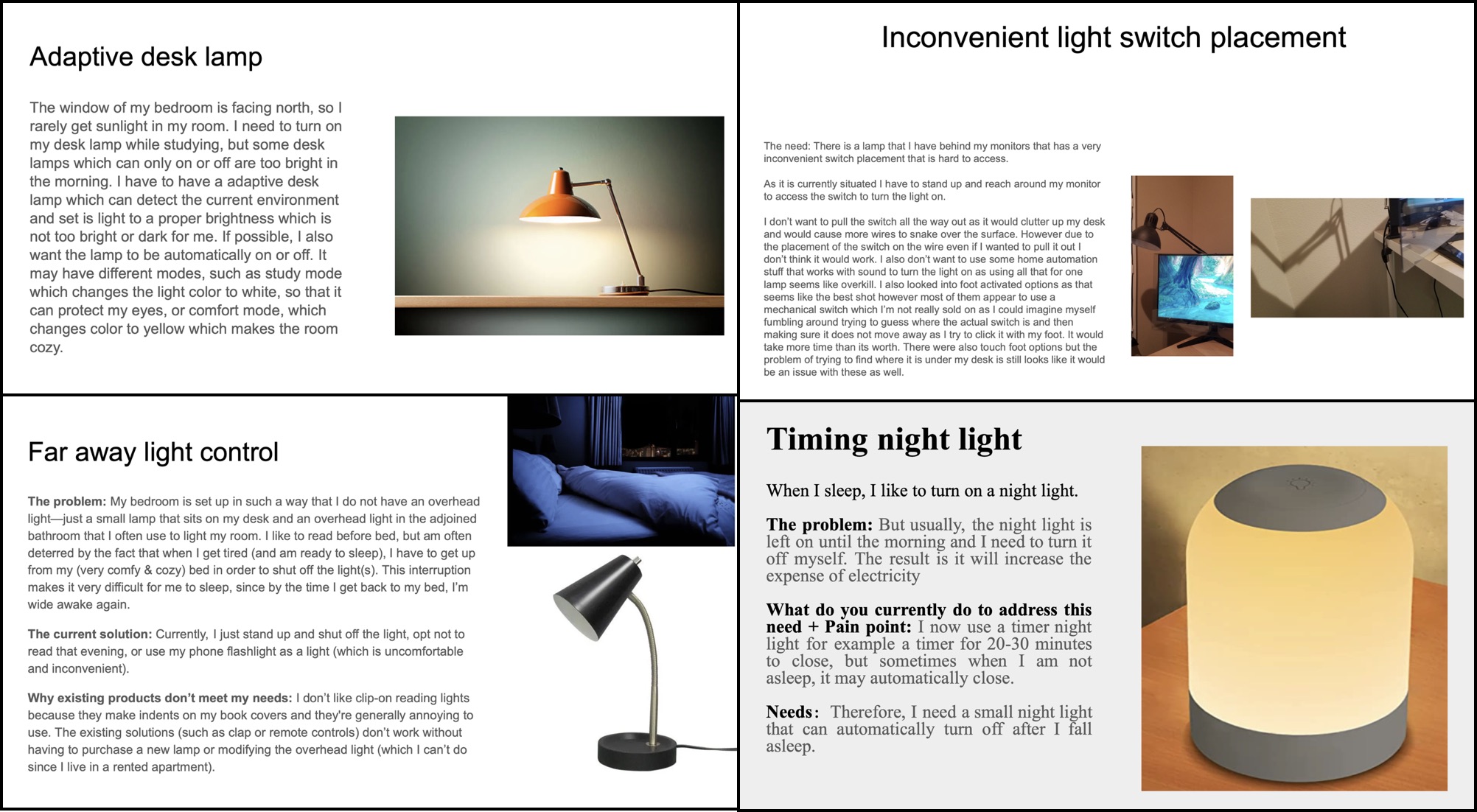}
    \caption{Variations of remote light controls.}
    \label{lights}
\end{figure}

\subsubsection{In-Class Activity Participation Score}
In-class activities (ICA) were mandatory, and accounted for 4\% of the total grade for the class. All the customer activities were assigned as ICAs. The first iteration of our class was offered in a quarter where it was hybrid instruction and thus led to some issues around attendance, and one team reported having trouble reaching their customer, who was online but did not respond. In the second iteration of our class, there was no reported issues and based on attendance, the customers (N=11) all showed up to ICAs. 

\subsubsection{Student dropping the class}
One worry our teaching team have is the situation where the peer-customer drops the class. This has yet to happen in the two sections offered. We believe this may be because the peer-customer selection process occurs after the initial add/drop period where most of the big enrollment shifts occur. A contingency plan that was made in case this occurred is to assign a TA to try to act as a stand-in customer. Although this is not a perfect solution, the TA can be expected to (1) be available in class and (2) apply a best effort judgement to give external feedback to the team on their prototype. 

\subsection{Idea Ownership}
In self-ideated projects, a common occurrence is a project team wants to continue working on the idea. Since they came up with the idea, they clearly own it. However, in this case, it is unclear who owns the idea. The peer-customer is personally invested because they expressed the desire to use it. The team, however, developed the solution. This has yet to occur. We suspect because the team is not so invested in the problem and treated it more like trying to deliver a solution for a client. Once delivered, they are done. Although this has yet to happen, we suspect in the future there may be a project that leads to this conflict. This will likely need to be resolved on a case-by-case basis. If the customer is interested in continuing to solve the problem and likes what the team has done so far, they can discuss with the team to keep working off of their solution. If the team wants to keep working on the solution, they may want to loop the customer into their design team. 

\section{Conclusion}
In this paper, we reported our experience using a new final project ideation mechanism called peer-customers. We found that peer-customers were effective at proposing interesting needs. More importantly, it was clearly observed in multiple cases that having a customer allowed for student teams to gain hands-on experience creating a solution based on real customer feedback. This solution can benefit from a number of improvements around more refined instructions provided during lectures, but even as it stands, did not present much issues around course logistics. We believe that peer-customers is a very effective way to provide a scalable mechanism for curating customers for student teams in hands-on curriculum. 
\bibliographystyle{ACM-Reference-Format}
\bibliography{references}


\end{document}